\documentclass[twocolumn,assymb,amsmath]{revtex4}
\usepackage{mathrsfs}
\usepackage{txfonts}
\usepackage{graphicx}
\usepackage{amssymb}
\usepackage{amsmath}
\usepackage{wasysym}
\usepackage{color}

\begin{document}

\title{Floquet theorem for open systems and its applications}
\author{C. M. Dai, Z. C. Shi,  X. X. Yi}
\affiliation{Center for Quantum Sciences and School of Physics,
Northeast Normal University, Changchun 130024, China}

\date{\today}

\begin{abstract}
For a closed system with periodic driving, Floquet theorem tells
that the time evolution operator can be written as $ U(t,0)\equiv
P(t)e^{\frac{-i}{\hbar}H_F t}$ with $P(t+T)=P(t)$, and $H_F$ is
Hermitian and time-independent  called Floquet Hamiltonian. In this
work, we extend the Floquet theorem from closed systems to open
systems described by a  Lindblad master equation that is periodic in time.
Lindbladian expansion  in powers of $\frac 1 \omega$ is derived, where
$\omega$ is the driving frequency.  Two examples are presented to
illustrate the theory. We find that appropriate trace preserving
time-independent Lindbladian of such a periodically driven system can
be constructed by the application of open system Floquet theory, and
it agrees well with the exact dynamics in the high frequency limit.
\end{abstract}

 \maketitle

\section{Introduction}
Periodically driven quantum dynamics  has recently attracted much
attention both in experiment and
theory\cite{deng15,eisert15,oka09,kitagawa10,lindner111,
lindner13,inoue101,rudner13,liu13,reynoso13,grushin14,thakurathi13,delplace13},
as it possesses novel properties such as topological
phases\cite{inoue100,lindner110,jiang11} and quantum phase
transitions\cite{ bastidas120,bastidas121} that otherwise  would be
impossible to achieve in the undriven case. Therefore, the
application of ac fields has become a very promising tool to
engineer quantum systems.

It is  difficult to handle a quantum system driven periodically by a
field with respect to its undriven counterpart, fortunately, Floquet
theorem  provides us with a method to deal with such a system. It
tells that for a closed system governed by a periodic Hamiltonian,
its evolution operator can be decomposed  into two parts, one can
given by  a time-independent effective Hamiltonian (called Floquet
Hamiltonian) and another  is periodic in time describing  the
periodic micromotion of the driven system(called micromotion
operator). Hence, we can design a suitable time-periodic driving to
change the properties of the Floquet Hamiltonian as well as  the
micromotion operator. This method has been used in many experiments
with ultracold atoms in driven optical lattices, for example, the
dynamic localization \cite{dunlap86,grossmann91,
holthaus92,grifoni98,lignier07,kierig08,eckardt09,creffield10}, the
control of the bosonic superfluid-to-Mott-insulator transition
\cite{eckardt05,zenesini09}, resonant coupling of Bloch bands
\cite{gemelke05,bakr11,parker13,ha15}, the dynamic creation of
kinetic frustration \cite{eckardt10,struck11}, the realization of
artificial magnetic fields and topological band structures
\cite{oka09,kolovsky11,bermudez11,aidelsburger11,struck12,hauke12,struck13,aidelsburger13,miyake13}
as well as the manipulation of spin-orbit coupling for cold
atoms\cite{jimnez15,luo15}.

For closed  systems,  periodically driven quantum dynamics is
governed  by a time-periodic Hamiltonian $H(t+T)=H(t)$ leading to an
unitary time-evolution operator
$U(t,0)=\mathscr{T}e^{-\frac{i}{\hbar}\int_{0}^{t}H(\tau)d\tau}$.
Because $H(t+T)=H(t)$, it is convenient to define a time-independent
Hamiltonian $H_F$ satisfying  $e^{-\frac{i}{\hbar}H_F T}=U(T,0)$. By
the use of $H_F$, we can rewrite $U(t,0)=U(t,0)
e^{\frac{i}{\hbar}H_F t} e^{-\frac{i}{\hbar}H_F t}$. Defining
$P(t)\equiv U(t,0) e^{\frac{i}{\hbar}H_F t}$, we can easily check
that $P(t+T)=P(t)$, where $P(t)$ is an operator describing
micromotion as mentioned above. $H_F$ is the effect Floquet
Hamiltonian. Having $H_F$,  we can use its eigenstate as a basis
$\{|\hbar\omega_{F}^{(n)}\rangle\}$. For any initial state, we can
expand it with these bases as
$|\Psi(0)\rangle=\sum_{n}c_{n}|\hbar\omega_{F}^{(n)}\rangle$, then
the state at time $t$ takes,
$|\Psi(t)\rangle=\sum_{n}c_{n}e^{-i\omega_{F}^{(n)}t}|\Phi(t)^{(n)}\rangle$,
where $|\Phi(t)^{(n)}\rangle=P(t)|\hbar\omega_{F}^{(n)}\rangle$
called Floquet modes. In the large frequency limit, we can consider
stroboscopic time evolution only, and the problem is then simplified to
calculate $H_F$. For some special cases, we can get an explicit
expression for $H_F$, but in general we can only get an
approximating result for $H_F$\cite{bukov15}.

In most realistic situations, a quantum system should be considered
as an open quantum system coupled to an environment that induces
decoherence and dissipation.  Such systems are of interest for
studies due to its connection to  quantum computation,  precision
measurements and theories of quantum measurement. However, a general
theory for open quantum systems similar to the Floquet theorem for
closed systems remains unexplored.

In this work, we present a Floquet theorem for open systems. We find
that the linear map which transforms the system from   initial
states  to  final states  can be decomposed into two parts. The
first part stems from the periodic time dependence of the driven
system and is called micromotion, while the second contribution,
which leads to deviations from the  periodic evolution, originates
from a time-independent Lindbladian. Using the Floquet theorem of
open system, we calculated the dynamics of a dissipative two-level
system with periodic Hamiltonian or Lindblad operators. We define a
fidelity to quantify the deviation of the exact dynamics to that by
the Floquet theorem,  and  find numerically the dependence of the
fidelity on the driven frequency and  decay rate etc..

\section{Formalism}
We start with a time dependent master  equation with time-dependent
Lindbladian $\mathscr{L}(t)$,
\begin{eqnarray}
\partial_t\rho(t)=\mathscr{L}(t)(\rho(t)).\label{master equation}
\end{eqnarray}
As $\mathscr{L}(t)$ is a linear operator, we  can define a linear
map $\mathscr{V}(t)$ by,
\begin{eqnarray}
\partial_t\mathscr{V}(t,t_1)=\mathscr{L}(t)\mathscr{V}(t,t_1),\label{trans equation}
\end{eqnarray}
so the solution of the Eq.(\ref{master equation}) is formally given
by,
\begin{eqnarray}
    \rho(t_2)=\mathscr{V}(t_2,t_1)(\rho(t_1)),\label{a1}
\end{eqnarray}
where the propagator $\mathscr{V}(t_2,t_1)$ takes,
$\mathscr{V}(t_2,t_1)= e^{\mathscr{L}(t_{2})\delta t_{n}}\ldots
e^{\mathscr{L}(t_{i})\delta t_{i}} \ldots
e^{\mathscr{L}(t_{1})\delta t_{1}}$, and by the divisibility
condition, we have
\begin{eqnarray}
    \mathscr{V}(t_2,t_1)=\mathscr{V}(t_2,t_0)\mathscr{V}(t_0,t_1).\label{a2}
\end{eqnarray}
If $\mathscr{L}(t)$ periodically depends on  time and the dynamics
is Markovian, $t_1$ and $t_2$ in $\mathscr{V}(t_2,t_1)$ are not
independent, i.e., the propagator  depends only on $t_2-t_1$. Noting
that at each time instance, there is a infinitesimal propagator,
$e^{\mathscr{L}(t_{i})\delta t_{i}}$  and
$\mathscr{L}(t)=\mathscr{L}(t+T)$, we can divide the time evolution
from $t_{1}$ to $t_{2}$ into three part, i.e., starting part, middle
part and  ending part. In order to find the three parts, we first
recall that,
\begin{eqnarray}
    \mathscr{V}(t_2,t_1)=\mathscr{V}(t_2+T,t_1+T),\label{a3}
\end{eqnarray}
then we can write the propagator in a compact form,
\begin{eqnarray}
&&\mathscr{V}(t_2,t_1)\nonumber\\
&&=\mathscr{V}(t_2,t_0+nT)\mathscr{V}(t_0+nT,t_0)\mathscr{V}(t_0,t_1),\nonumber\\
&&=\mathscr{V}(t_2,t_0+nT)e^{n\mathscr{L}_{F}[t_0]T}\mathscr{V}(t_0,t_1),\\
&&=\mathscr{V}(t_2,t_0+nT)e^{-\mathscr{L}_{F}[t_0]\delta t_2}
e^{\mathscr{L}_{F}[t_0](t_2-t_1)}e^{\mathscr{L}_{F}[t_0]\delta t_1}\mathscr{V}(t_0,t_1),\nonumber\\
&&=\mathscr{K}(\delta
t_2)e^{\mathscr{L}_{F}[t_0](t_2-t_1)}\mathscr{J}(\delta
t_1),\nonumber\label{a4}
\end{eqnarray}
where
$\mathscr{K}(t)=\mathscr{V}(t_0+t,t_0)e^{-\mathscr{L}_{F}[t_0]t}$,
$\mathscr{J}(t)=e^{\mathscr{L}_{F}[t_0]t}\mathscr{V}(t_0,t_0+t)$,
$\mathscr{V}(t_0+T,t_0)=e^{\mathscr{L}_{F}[t_0]T}$, $\delta
t_2=t_2-(nT+t_0)$ and  $\delta t_1=t_1-t_0$. $\mathscr{L}_{F}[t_0]$
will be referred  to  effective generator in later discussion. The
form of $\mathscr{L}_{F}[t_0]$ depends on the algebraic structure of
$\mathscr{L}(t)$. We use the argument  $t_{0}$ to denote  the
dependence of  $\mathscr{L}_{F}$  on the starting  time $t_0$.
Different starting time corresponds to different
$\mathscr{V}(t_0+T,t_0)$. Thus for each $t_{0}$,  we would have a
set
$\mathscr{L}_{F-ALL}[t_0]=\{\mathscr{L}_{F}[t_0]|\mathscr{V}(t_0+T,t_0)=e^{\mathscr{L}_{F}[t_0]T}\}$.
For different $t_0$, the set is different.

In practice, to study the dynamics of an open system, we do not need to find all sets
$\mathscr{L}_{F-ALL}[t_0]$: one element  in
$\mathscr{L}_{F-ALL}[t_0]$ is fine. For a time-dependent generator
$\mathscr{L}(t)$,  Magnus proposed a method to find the approximate
solution for $\mathscr{L}_{F}$ \cite{kitagawa10,blanes09}, a high-frequency
expansion for $\mathscr{L}_{F}$ can be found using the
Baker-Campbell-Hausdroff lemma\cite{bukov15}, giving the middle part mentioned earlier.

As to the other two parts in the propagator, we can verify that
$\mathscr{K}(t)$ and $\mathscr{J}(t)$ periodically depend on time,
namely
\begin{eqnarray}
    &&\mathscr{K}(t+T)\nonumber\\
    &&=\mathscr{V}(t_0+t+T,t_0)e^{-\mathscr{L}_{F}[t_0](t+T)},\nonumber\\
    &&=\mathscr{V}(t_0+t+T,t_0+T)\mathscr{V}(t_0+T,t_0)e^{-\mathscr{L}_{F}[t_0](t+T)},\\
    &&=\mathscr{K}(t),\nonumber\label{a5}
\end{eqnarray}
for $\mathscr{J}(t)$ the same proof works. By the definition of
$\mathscr{K}(t)$ and $\mathscr{J}(t)$, it is easy to find that
\begin{eqnarray}
\partial_t\mathscr{K}(t)=\mathscr{L}(t+t_0)\mathscr{K}(t)-\mathscr{K}(t)\mathscr{L}_{F}[t_0],\label{K equation}
\end{eqnarray}
\begin{eqnarray}
\partial_t\mathscr{J}(t)=\mathscr{L}_{F}[t_0]\mathscr{J}(t)-\mathscr{J}(t)\mathscr{L}(t+t_0),\label{J equation}
\end{eqnarray}
Clearly the form of  $\mathscr{K}(t)$ and $\mathscr{J}(t)$ depends
on the choice of $t_0$ and $\mathscr{L}_{F}[t_0]$.

When we choose $t_0$ and $\mathscr{L}_{F}[t_0]$ to satisfy,
\begin{eqnarray}
\mathscr{V}(t_0+T,t_0)=e^{\mathscr{L}_{F}[t_0]T},\label{LF equation}
\end{eqnarray}
the starting and ending parts, i.e., $\mathscr{K}(t)$ and $\mathscr{J}(t)$ can be established. By the
definition of  $\mathscr{L}_{F}$ in Eq. (\ref{LF equation}), we find
that $\mathscr{L}_{F}$ itself can give a propagator for the
evolution time $T$. In other words, given an initial state (at time
$t_0$) of an open system, $\mathscr{L}_{F}$ itself  can map the
initial state to the final state at time $t_0+T$. One may wonder,
how can we know the state of the system at a middle time, say
$t_0<t<t_0+T$? Eq. (\ref{a4}) shows that $\mathscr{K}(t)$ and
$\mathscr{J}(t)$ would help.

Now we show how to calculate $\mathscr{L}_{F}$.  Without loss of
generality, we set $t_1=t_0=0$,  so $\mathscr{J}(\delta t_1)=1$. By
the use of Magnus expansion, we can derive an expression for
$\mathscr{L}_{F}$ from Eq.(\ref{LF equation}). This method is
available at high driving frequency, but it breaks down at low
frequency. To find a $\mathscr{L}_{F}$ satisfying Eq.(\ref{LF
equation}), we write
\begin{eqnarray}
\mathscr{V}(t)=e^{\Phi(t)}.\label{b1}
\end{eqnarray}
Obviously, $\mathscr{L}_{F}=\Phi(T)/T$  satisfies Eq.(\ref{LF
equation}).  We can obtain an expansion  for $\Phi(T)$  by Magnus
expansion,
\begin{eqnarray}
\Phi(T)=\sum_{n=0}^{\infty}{\Phi^{(n)}(T)}.\label{b2}
\end{eqnarray}
Similarly, $\mathscr{L}_{F}^{(n)}=\Phi^{(n)}(T)/T$.  The first three
terms are,
\begin{eqnarray}
&&\mathscr{L}_{F}^{(0)}=1/T\int_{0}^{T}\mathscr{L}(t)dt,\nonumber\\
&&\mathscr{L}_{F}^{(1)}=1/(2T)\int_{0}^{T}dt_{1}\int_{0}^{t_1}dt_{2} [\mathscr{L}(t_{1}),\mathscr{L}(t_{2})],\\
&&\mathscr{L}_{F}^{(2)}=1/(6T)\int_{0}^{T}dt_{1}\int_{0}^{t_1}dt_{2}\int_{0}^{t_2}dt_{3}\nonumber\\
&&\{[\mathscr{L}(t_{1}),[\mathscr{L}(t_{2}),\mathscr{L}(t_{3})]]+ [[\mathscr{L}(t_{1}),\mathscr{L}(t_{2})],\mathscr{L}(t_{3})]\},\nonumber\label{b3}
\end{eqnarray}
High-order  terms can be obtained by  recursion not presented here.
Because $\mathscr{K}(t)$ and $\mathscr{L}(t)$   periodically depend
on time,  we can expand  them into Fourier series,
\begin{eqnarray}
\mathscr{K}(t)=\sum_{m=-\infty}^{\infty}\mathscr{K}_{m}e^{i\omega mt},\label{K series}
\end{eqnarray}
\begin{eqnarray}
\mathscr{L}(t)=\sum_{m=-\infty}^{\infty}\mathscr{L}_{m}e^{i\omega
mt}.\label{L series}
\end{eqnarray}
Substituting these equations into  Eq.(\ref{K equation}), we find
\begin{eqnarray}
i\omega m\mathscr{K}_{m}=\sum_{n}\mathscr{L}_{n}\mathscr{K}_{m-n}
-\mathscr{K}_{m}\mathscr{L}_{F}.\label{series equation}
\end{eqnarray}
Eq.(\ref{K series}) and Eq.(\ref{L series}) need to be truncated in
order to solve Eq.(\ref{series equation}).

Notice that  $\mathscr{L}$ would drive  any initial states into a
Floquet steady state defined by
$\rho_{F}=e^{\mathscr{L}_{F}t}(\rho(0))$, though sometimes we can
not use Magnus expansion to get $\mathscr{L}_{F}$, but we always
know $\mathscr{L}_{F}(\rho_{F})=0$ with $t\rightarrow\infty$. So
using Eq.(\ref{series equation}), we obtain
\begin{eqnarray}
i\omega m\rho_{m}=\sum_{n}\mathscr{L}_{n}(\rho_{m-n}),\label{rhos equation}
\end{eqnarray}
where $\rho_{m}$ is the Fourier coefficients of final state, i.e.
\begin{eqnarray}
\left[
\begin{matrix}
\ddots&\ddots&\ddots&\ddots&\ddots&\\
\ddots&\mathscr{L}_{0}+i\omega&\mathscr{L}_{-1}&\mathscr{L}_{-2}&\ddots&\\
\ddots&\mathscr{L}_{1}&\mathscr{L}_{0}&\mathscr{L}_{-1}&\ddots&\\
\ddots&\mathscr{L}_{2}&\mathscr{L}_{1}&\mathscr{L}_{0}-i\omega&\ddots&\\
\ddots&\ddots&\ddots&\ddots&\ddots&
\end{matrix}
\right]\left[
\begin{matrix}
\vdots\\
\rho_{-1}\\
\rho_{0}\\
\rho_{1}\\
\vdots
\end{matrix}
\right] = 0,\label{rhos martix}
\end{eqnarray}
For closed systems, the time-dependent Lindbladian takes
$\mathscr{L}(t)(\cdot)=-i[H(t),(\cdot)]$, Eq.(\ref{master equation})
then returns  to  the quantum Liouville equation
$\partial_t\rho(t)=-i[H(t),\rho(t)]$, where $H(t)$ is the system
Hamiltonian. In this case, consider an infinitesimal time $\delta
t$, $e^{\mathscr{L}(t)\delta t}(\cdot)$ can be written as
$e^{-iH(t)\delta t}(\cdot) e^{iH(t)\delta t}$. So
$\mathscr{V}(t,0)(\cdot)=e^{-iH(t)\delta t}...e^{-iH(t_{i})\delta
t}...e^{-iH(0)\delta t} (\cdot) e^{iH(0)\delta
t}...e^{iH(t_{i})\delta t}...e^{iH(t)\delta
t}=U(t,0)(\cdot)U^{\dag}(t,0)$, where $U(t,0)$ is an unitary
time-evolution operator. By the spirit mentioned,  we can choose
$\mathscr{L}_{F}$ in the form $-i[H_{F},(\cdot)]$ to satisfy
Eq.(\ref{LF equation}). If $e^{-iH_{F}T}=U(T,0)$,
$e^{iH_{F}T}=U^{\dag}(T,0)$ is naturally satisfied. Note that
$H_{F}$ is Hermitian\cite{rahav03}, because $U^{\dag}$ in unitary,
we have $\mathscr{K}(t)(\cdot)=P(t)(\cdot)P^{\dag}(t)$. These
observations together lead to
$\mathscr{V}(t,0)(\cdot)=\mathscr{K}(t)e^{\mathscr{L}_{F}t}(\cdot)=P(t)e^{-iH_{F}t}(\cdot)
e^{iH_{F}t} P^{\dag}(t)$, covering the Floquet theorem for  closed
quantum systems.

\section{Example}
In this section we illustrate our  theory with two examples.  Both
of them describe  a two-level system subject to decoherence. In the
first example, the Lindblad operator is a periodic function of time,
whereas in the second the Hamiltonian is periodic in time. The
master equation that describes the first example takes,
\begin{eqnarray}
\partial_t\rho(t)=-i[H,\rho(t)]+\mathscr{D}(t)(\rho(t)),\label{model1 equation}
\end{eqnarray}
in which
$\mathscr{D}(t)(\rho(t))=\gamma(2A(t)\rho(t)A^{\dagger}(t)-\{A^{\dagger}(t)A(t),\rho(t)\})$
and $H=\Omega\sigma_z$, and $A(t)=cos(\omega t)\sigma_{+}+sin(\omega
t)\sigma_{-}$. This type of master equation is derived in \cite{kamleitner11}
that can be used to describe Cooper-pair pumping. By the use of Magnus
expansion, we  work out the  first two leading terms  of
$\mathscr{L}_{F}$ in large frequency limit,
\begin{eqnarray}
{\mathscr{L}_{F}}^{(0)}(\cdot)=-i[H,\cdot]+
\gamma(\sigma_{+}\cdot\sigma_{-}+\sigma_{-}\cdot\sigma_{+}-I(\cdot)),\label{1LF0}
\end{eqnarray}
\begin{eqnarray}
{\mathscr{L}_{F}}^{(1)}(\cdot)=
2i\gamma(\Omega/\omega)(\sigma_{-}\cdot\sigma_{-}-\sigma_{+}\cdot\sigma_{+}),\label{1LF1}
\end{eqnarray}
and $\mathscr{L}_{F}=\mathscr{L}_{F}=
{\mathscr{L}_{F}}^{(0)}+{\mathscr{L}_{F}}^{(1)}+\mathcal{O}(1/\omega^2).$

\begin{figure}
\includegraphics[scale=0.28]{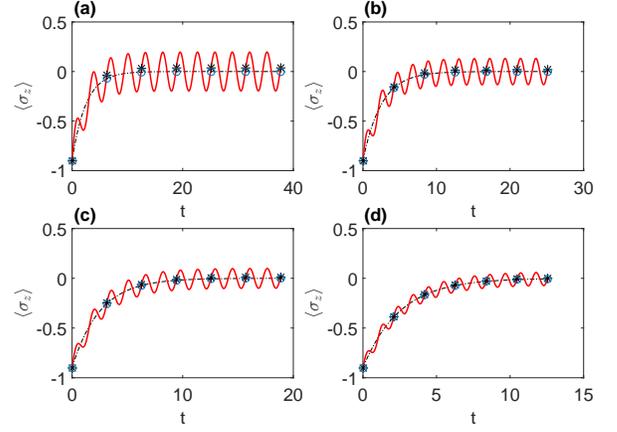}
\caption{(Color online)  $\langle\sigma_z\rangle$ as a function of
time. Parameters $\gamma=0.2$, $\Omega=1$ are chosen in each figure.
(a)-(d) are for different driving frequency: (a) $\omega=\Omega$,
(b) $\omega=1.5\Omega$, (c) $\omega=2\Omega$, and (d)
$\omega=3\Omega$. Red line is numerical calculation with exact
$\mathscr{L}(t)$. $\langle\sigma_z\rangle$ at the  period points of
time  are marked by $\hexstar$ on the line. Blue circles mark the
result obtained  by the approximate  $\mathscr{L}_{F} \approx
{\mathscr{L}_{F}}^{(0)}+{\mathscr{L}_{F}}^{(1)}$. The results  show
that  as frequency increases the amplitude of oscillations
decreases, and the approximate   $\mathscr{L}_{F}$  gets  closer to
the exact $\mathscr{L}_{F}$.} \label{fig1}
\end{figure}

\begin{figure}
\includegraphics[scale=0.28]{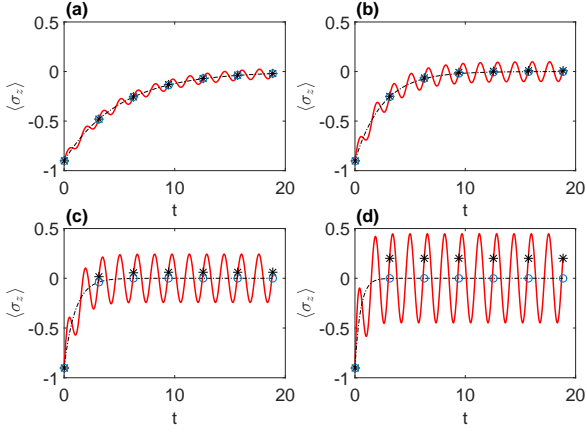}
\caption{(Color online) $\langle\sigma_z\rangle$ as a function of
time. The lines are plotted  with $\omega=2\Omega$, $\Omega=1$.
(a)-(d) are for different $\gamma$, namely (a) $\gamma=0.1$,
(b)$\gamma=0.2$, (c)$\gamma=0.5$, and  (d)$\gamma=1$. Red line is
the numerical result with exact $\mathscr{L}(t)$. The value at the
period points in time  are marked by \hexstar.  Blue circles mark
the result numerically obtained  by the approximate $\mathscr{L}_{F}
\approx {\mathscr{L}_{F}}^{(0)}+{\mathscr{L}_{F}}^{(1)}$. Because
terms ${\mathscr{L}_{F}}^{(n)}$ that are  high order in $\frac
1\omega$ are not zero and depend on $\gamma$. So for larger
$\gamma$, $\mathscr{L}_{F}-
({\mathscr{L}_{F}}^{(0)}+{\mathscr{L}_{F}}^{(1)})$
 no more approaches zero, as shown  by the deviation  of blue circle from
the $\hexstar$.}\label{fig2}
\end{figure}

Fig. \ref{fig1} and  Fig. \ref{fig2} show the difference  between
the averages of $\sigma_z$ given by
${\mathscr{L}_{F}}^{(0)}+{\mathscr{L}_{F}}^{(1)}$ and by the exact
$\mathscr{L}_{F}$.
$\circ$ markers  the periodic points of time. Here the period is
chosen to be $T=2\pi/\omega$, then $\mathscr{L}(t)$ has period
$T/2$. Fig.\ref{fig1} shows that  the amplitude of oscillation of
$\langle\sigma_z\rangle$ is reduced as  frequency increases. From
Fig.\ref{fig2} we find that as  $\gamma$ increases, the amplitude of
oscillation is enhanced, although for a time-independent open
system,  $\gamma$ is the decay rate.

These results suggest that the Magnus expansions can correctly
predict the asymptotical behavior of evolution, but for large
$\gamma$, high-order  Mangus terms should be included to get more
accurate results.

The relationship between the amplitude of oscillation and the
frequency might  be obtained  from Eq.(\ref{rhos martix}). But in
the high  frequency limit, we can use Eq.(\ref{K equation}) to get
an well  result. Indeed,  for evolution time in $[0, T]$, we have
the Mangus expansion $\mathscr{K}(t)=
e^{\Sigma_{n=1}^{\infty}\Lambda^{(n)}(t)}$. Note that for every $n$,
 $\Lambda^{(n)}(t)$ preserves the periodicity. In fact, we can write
$\Lambda^{(n)}(t)=\frac{1}{\omega^n}\mathscr{A}^{(n)}(t)$, where
$\mathscr{A}^{(n)}(t)$ depends on the Fourier coefficient of
$\mathscr{L}(t)$ \cite{blanes09,bukov15}. Then we have $\mathscr{K}(t)\approx
1+\frac{1}{\omega}\mathscr{A}^{(1)}(t)$ and
$\frac{1}{\omega}\mathscr{A}^{(1)}(t)=\frac{1}{\omega}(\int_0^{t}\mathscr{L}(\tau)d\omega\tau-\omega
t\mathscr{L}^{(0)}_{F})$. When $\mathscr{L}(t)=\mathscr{F}(\omega
t,\alpha,\beta, \cdots)$, the amplitude of oscillation $\sim
\frac{1}{\omega}$.


In the second model, we consider a spin-$\frac 1 2 $ in a
time-dependent magnetic field. The master equation is then,
\begin{eqnarray}
\partial_t\rho(t)=-i[H(t),\rho(t)]+\mathscr{D}(\rho(t)),\label{model2 equation}
\end{eqnarray}
with  $H(t)=1/2 \alpha
\vec{B}(t)\vec{\sigma}$,$\vec{B}(t+T)=\vec{B}(t)$ and
$\mathscr{D}(\rho(t))=\gamma(2\sigma_{-}\rho(t)
\sigma_{+}-\sigma_{+}\sigma_{-}\rho(t)-\rho(t)\sigma_{+}\sigma_{-})$.
The  first two leading terms of $\mathscr{L}_{F}$ are,
\begin{eqnarray}
{\mathscr{L}_{F}}^{(0)}(\cdot)=-i[\overline{H(t)},\cdot]+\mathscr{D}(\cdot),\label{2LF0}
\end{eqnarray}
\begin{eqnarray}
&&{\mathscr{L}_{F}}^{(1)}(\cdot)=(i/2)\alpha^{2}[M_{x}\sigma_{x}+M_{y}\sigma_{y}+M_{z}\sigma_{z},\cdot]\nonumber\\
&&+(i/2)\alpha\gamma(N_{x}+iN_{y})(2\sigma_{-}\cdot\sigma_{z}+\{\cdot,\sigma_{-}\})\\
&&-(i/2)\alpha\gamma(N_{x}-iN_{y})(2\sigma_{z}\cdot\sigma_{+}+\{\cdot,\sigma_{+}\}),\nonumber\label{2LF1}
\end{eqnarray}
where
\begin{eqnarray}
&&M_{x}=\overline{\overline{B_{y}(t_2)B_{z}(t_1)-B_{y}(t_1)B_{z}(t_2)}}\nonumber\\
&&M_{y}=\overline{\overline{B_{z}(t_2)B_{x}(t_1)-B_{z}(t_1)B_{x}(t_2)}}\nonumber\\
&&M_{z}=\overline{\overline{B_{x}(t_2)B_{y}(t_1)-B_{x}(t_1)B_{y}(t_2)}},\nonumber\label{M}
\end{eqnarray}
and
\begin{eqnarray}
&&N_{x}=\overline{\overline{B_{x}(t_1)-B_{x}(t_2)}}\nonumber\\
&&N_{y}=\overline{\overline{B_{y}(t_1)-B_{y}(t_2)}},\nonumber\label{N}
\end{eqnarray}
here $\overline{\overline{f(t_{1},t_{2})}}=1/(2T)\int_{0}^{T}dt_{1}\int_{0}^{t_1}dt_{2}f(t_1,t_2)$.

If we consider $\vec{B}$ rotates around an axis with spherical
coordinate $(\theta,\varphi)$, and the angle of deviation from the
z-axis  is $\beta$, i.e. $\vec{B}(t)=\vec{B}_p+\vec{B}_v(t)$,
$\vec{B}_p=\cos(\beta)(\cos(\theta)\sin(\varphi),
\sin(\theta)\sin(\varphi), \cos(\varphi))$ and $\vec{B}_v(t)=\sin
(\beta)(\sin(\omega t)(\cos(\theta)\cos(\varphi), \sin(\theta)\cos
(\varphi), -\sin(\varphi))+\cos(\omega t)(\sin(\theta),
-\cos(\theta), 0))$, we have,
\begin{eqnarray}
&&M_{x}=1/2\omega\sin\beta(\sin\beta\cos\theta\sin\varphi+2\cos\beta\sin\theta)\nonumber\\
&&M_{y}=1/2\omega\sin\beta(\sin\beta\sin\theta\sin\varphi-2\cos\beta\cos\theta)\nonumber\\
&&M_{z}=1/2\omega\cos\varphi\sin^{2}\beta\nonumber\\
&&N_{x}=-1/\omega\cos\theta\cos\varphi\sin\beta\nonumber\\
&&N_{y}=-1/\omega\sin\theta\cos\varphi\sin\beta\nonumber\\
&&\overline{H(t)}=(1/2)\alpha\cos\beta[\cos\varphi\sigma_{z}
+\sin\varphi(\cos\theta\sigma_{x}+\sin\theta\sigma_{y})].\nonumber\label{Mrot}
\end{eqnarray}

\begin{figure}
\includegraphics[scale=0.28]{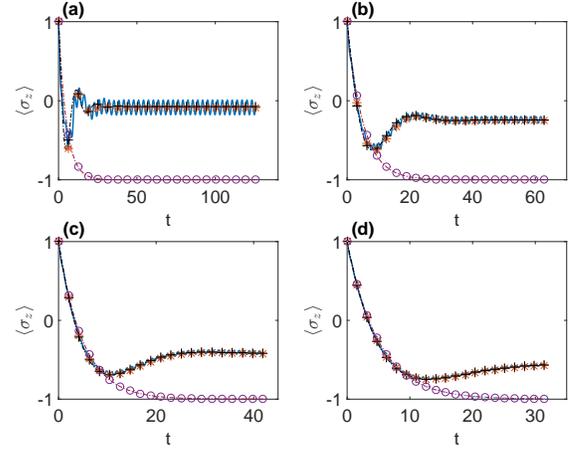}
\caption{(Color online) The average of $\sigma_z$ versus time.
Parameter  $\theta=\frac{\pi}{4}$, $\varphi=\frac{\pi}{2}$,
$\beta=\frac{\pi}{2}$, $\alpha=1$, and  $\gamma=0.1$ are chosen for
these lines.  $\omega=1,2,3,4$ are  for (a), (b), (c), and (d),
respectively.  Blue line is  the numerical result by exact
$\mathscr{L}(t)$. The values at the period points in time  are
marked by \hexstar. Purple circles mark the result gotten  by the
approximate   $\mathscr{L}_{F}$ up to zeroth order. $+$ marks the
result obtained  by the approximate  $\mathscr{L}_{F}$ up to the
order in the expansion of $\mathscr{L}(t)$. }\label{fig3}
\end{figure}

\begin{figure}
\includegraphics[scale=0.28]{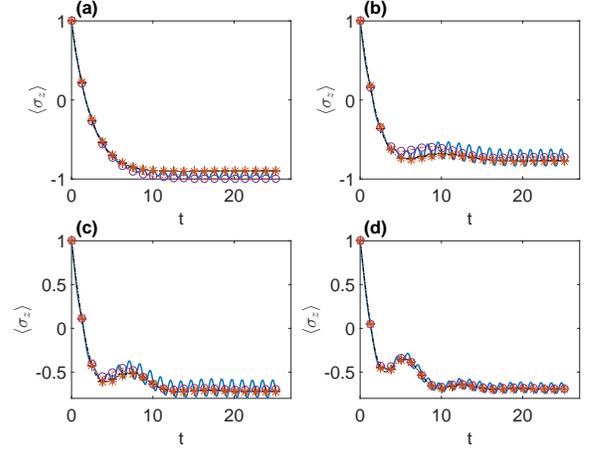}
\caption{(Color online)  The average of $\sigma_z$ versus time.
Parameters chosen are $\omega=5$, $\theta=\frac{\pi}{4}$,
$\varphi=\frac{\pi}{4}$, $\alpha=1$, and $\gamma=0.2$.
$\beta=\frac{\pi}{2},\frac{\pi}{3},\frac{\pi}{4},$ and
$\frac{\pi}{8}$ for (a), (b), (c), and  (d), respectively. Blue line
is the numerical calculation with exact $\mathscr{L}(t)$. The period
points in time  are marked by \hexstar. Purple circles mark the
result by the approximate  $\mathscr{L}_{F}$ up to the zeroth order
(first term only) in the expansion.  $+$ marks the result  by the
approximate $\mathscr{L}_{F}$ up to the first order (first term and
second term) in the expansion.}\label{fig4}
\end{figure}

As Eq. (\ref{2LF1}) shows, the first order of the effective
generator includes three parts. The first part behaves like  an
effective Hamilton $\frac{1}{2} \alpha^2 \textbf{M}
\textbf{$\sigma$}$. And \textbf{M} is a function of
$\theta,\varphi,\beta,\omega$,
\textbf{M}=$\frac{1}{\omega}\textbf{m}(\theta,\varphi,\beta)$. The
second and third terms are inversely proportional to
$\frac{1}{\omega}$. For a  special case of
$\theta=\frac{\pi}{4}$,$\varphi=\frac{\pi}{2}$ and
$\beta=\frac{\pi}{2}$, terms with any one of
$\overline{H(t)}$,$N_{x}$,$N_{y}$ or $M_{z}$ vanish. In this case if
$\gamma$ is smaller  than $\alpha^2$,
${\mathscr{L}_{F}}^{(1)}(\cdot)$ would dominate the dynamics. To
this extend, we can say the system is sensitive to the frequency of
the driving $B$, because ${\mathscr{L}_{F}}^{(1)}(\cdot) \sim
1/\omega$. See   Fig.\ref{fig3} and  Fig.\ref{fig4}.

\section{Conclusion and discussions}
We have derived a general approach to solve periodically driven
systems subject to decoherence, which we call the Floquet theorem of
open system. The theorem  allows to obtain an effective
time-independent Lindbladian for  different driving regimes. We show
that in the high-frequency limit, the leading-order of the Mangus
expansion agrees well with the exact dynamics. When we generalize
the results to an open system  with two periods, say period $T_1$
and $T_2$, higher order expansion in $\frac 1\omega$ than a system
with only $T_1$ or $T_2$ periodicity   is  necessary. The reason is
as follows. Consider $T_1/T_2\approx p/q$, $p,q$ are prime number,
the overall period of  the  system  becomes $T_2 lcm(p,q)/q$ ( with
$lcm$ denoting the  lowest common multiple), which usually is  much
bigger than $T_1$ and $T_2$. So, higher order Mangus expansion has
to be taken into account.

\section*{ACKNOWLEDGMENTS}
This work is supported by National Natural Science Foundation of
China (NSFC) under Grants No. 11175032, and No. 61475033.

\end{document}